\documentclass[aps,prd,nofootinbib,amsmath,amssymb,superscriptaddress,twocolumn,10pt]{revtex4}%superscriptaddress,showpacs,

\pdfoutput=1

\usepackage{graphicx}
\usepackage{dcolumn}
\usepackage{bm}
\usepackage{amssymb}
\usepackage{latexsym}
\usepackage{booktabs}
\usepackage{amsmath}
\usepackage{multirow}
\usepackage[colorlinks=true, linkcolor=red, citecolor=blue]{hyperref}
\usepackage{mathrsfs}

\newcommand{\be}{\begin{equation}}
\newcommand{\ee}{\end{equation}}
\newcommand{\bq}{\begin{eqnarray}}
\newcommand{\eq}{\end{eqnarray}}

\begin{document}

\title{Exploring neutrino mass and mass hierarchy in the scenario of vacuum energy interacting with cold dark matter}

\author{Rui-Yun Guo}
\affiliation{Department of Physics, College of Sciences,
Northeastern University, Shenyang 110819, China}
\author{Jing-Fei Zhang }
\affiliation{Department of Physics, College of Sciences,
Northeastern University, Shenyang 110819, China}
\author{Xin Zhang\footnote{Corresponding author}}
\email{zhangxin@mail.neu.edu.cn}
\affiliation{Department of Physics, College of Sciences,
Northeastern University, Shenyang 110819, China}
\affiliation{Center for High Energy Physics, Peking University, Beijing 100080, China}

\begin{abstract}
We investigate the constraints on total neutrino mass in the
scenario of vacuum energy interacting with cold dark matter. We
focus on two typical interaction forms, i.e., $Q=\beta H\rho_{\rm
c}$ and $Q=\beta H\rho_{\Lambda}$. To avoid the occurrence of
large-scale instability in interacting dark energy cosmology, we
adopt the parameterized post-Friedmann approach to calculate the
perturbation evolution of dark energy. We employ observational
data, including the Planck cosmic microwave background temperature
and polarization data, baryon acoustic oscillation data, a
JLA sample of type Ia supernovae observation, direct
measurement of the Hubble constant, and redshift space
distortion data. We find that, compared with those in the
$\Lambda$CDM model, much looser constraints on $\sum m_{\nu}$ are
obtained in the $Q=\beta H\rho_{\rm c}$ model, whereas slightly
tighter constraints are obtained in the $Q=\beta H\rho_{\Lambda}$
model. Consideration of the possible mass hierarchies of neutrinos reveals that the
smallest upper limit of $\sum m_{\nu}$ appears in the
degenerate hierarchy case. By comparing the values of $\chi^2_{\rm
min}$, we find that the normal hierarchy case is favored over
the inverted one. In particular, we find that the difference
$\Delta \chi^2_{\rm min} \equiv \chi^2_{\rm IH; min}-\chi^2_{\rm
NH; min}> 2$ in the $Q=\beta H\rho_{\rm c}$ model. In addition, we
find that $\beta=0$ is consistent with the current observations in
the $Q=\beta H\rho_{\rm c}$ model, and $\beta < 0$ is favored at
more than the $1\sigma$ level in the $Q=\beta H\rho_{\Lambda}$
model.
\end{abstract}

\maketitle

\section{Introduction}

The phenomenon of neutrino oscillation has proved that neutrinos
have masses and that there are mass splittings between different
neutrino species (see
Refs.~\cite{Lesgourgues:2006nd,Agashe:2014kda} for reviews).
However, it is enormously difficult for particle physics
experiments to directly measure the absolute masses of neutrinos.
In fact, the solar and reactor experiments measured the squared
mass difference, $\Delta m_{21}^{2}\simeq 7.5\times 10^{-5}$
eV$^{2}$, and the atmospheric and accelerator beam experiments
measured the squared mass difference, $|\Delta m_{31}^{2}|\simeq
2.5\times 10^{-3}$ eV$^{2}$~\cite{Agashe:2014kda}. Thus, two
possible mass hierarchies are obtained, i.e., the normal hierarchy
(NH) with $m_{1}<m_{2}\ll m_{3}$ and the inverted hierarchy (IH)
with $m_{3}\ll m_{1}< m_{2}$, where $m_{i}$ ($i=1,~2,~3$) denotes
the masses of neutrinos in the three mass eigenstates. If the
mass splittings are neglected, we then have
$m_{1}=m_{2}=m_{3}$, which represents the degenerate hierarchy
(DH).

Actually, the absolute masses of neutrinos could in principle be
measured by particle physics experiments, such as tritium
beta decay
experiments~\cite{Osipowicz:2001sq,Kraus:2004zw,Otten:2008zz,Wolf:2008hf}
and neutrinoless double beta decay (0$\nu \beta \beta$)
experiments~\cite{KlapdorKleingrothaus:2002ip,KlapdorKleingrothaus:2004wj}.
In addition, experiments for detecting cosmic relic neutrinos
(e.g., the PTOLEMY
proposal~\cite{Betts:2013uya,Zhang:2015wua,Huang:2016qmh,Zhang:2017ljh})
are also able to measure the absolute masses of neutrinos.
However, compared with these particle physics experiments,
cosmological observations are considered to be a more promising
method to determine the absolute masses of neutrinos. Massive
neutrinos can leave rich signatures on the cosmic microwave
background (CMB) anisotropies and the large-scale structure (LSS)
formation in the evolution of the universe. Thus, one might extract
useful information on neutrinos from these available cosmological
observations.

Recently, the constraint on the total neutrino mass has been
reduced to $\sum m_{\nu}<0.15$ eV ($2\sigma$)~\cite{Zhao:2016ecj}
in the base $\Lambda$ cold dark matter ($\Lambda$CDM) model. For
dynamical dark energy models, the constraints become $\sum
m_{\nu}<0.25$ eV ($2\sigma$) in the $w$CDM model and $\sum
m_{\nu}<0.51$ eV ($2\sigma$) in the Chevallier--Polarski--Linder
(CPL) model~\cite{Zhao:2016ecj}, indicating that larger neutrino
masses are favored in the two dynamical dark energy models. However,
in the holographic dark energy (HDE) model, the constraint result
is reduced to $\sum m_{\nu}<0.113$ eV
($2\sigma$)~\cite{Zhang:2015uhk}, which is close to the edge of
diagnosing the mass hierarchy of neutrinos. For more studies on
constraining the total neutrino mass in cosmological models, see,
e.g.,
Refs.~\cite{Hu:1997mj,Reid:2009nq,Thomas:2009ae,Carbone:2010ik,Li:2012vn,Wang:2012uf,Audren:2012vy,Riemer-Sorensen:2013jsa,Font-Ribera:2013rwa,Zhang:2014dxk,Zhang:2014nta,Zhang:2014ifa,Palanque-Delabrouille:2014jca,Geng:2014yoa,Li:2015poa,Ade:2015xua,Zhang:2015rha,Geng:2015haa,Chen:2015oga,Allison:2015qca,Cuesta:2015iho,Chen:2016eyp,Moresco:2016nqq,Lu:2016hsd,Kumar:2016zpg,Xu:2016ddc,Vagnozzi:2017ovm,Zhang:2017rbg,Lorenz:2017fgo,Zhao:2017jma,Vagnozzi:2018jhn,Wang:2018lun}.

Furthermore, when the possible mass hierarchies of neutrinos are
considered, some previous studies showed that the NH case fits
cosmological observations better than the IH case. For example,
Huang et al.~\cite{Huang:2015wrx} gave the result $\Delta
\chi^{2}\equiv \chi^{2}_{\rm IH; min}-\chi^{2}_{\rm NH; min}
\simeq 3.38$ in the $\Lambda$CDM model; Wang et
al.~\cite{Wang:2016tsz} gave the results $\Delta \chi^{2}
\simeq 2.1$ in the $w$CDM model and $\Delta \chi^{2} \simeq 4.1$
in the HDE model. Further, when the DH
case is included for comparison, the DH case fits almost all
of the data combinations best. Consistent conclusions are also obtained
in the CPL model~\cite{Yang:2017amu}.

In addition, when interaction between dark energy and dark
matter is considered in current cosmology, the constraint on $\sum
m_{\nu}$ is reduced to $\sum m_{\nu}<0.10$ eV
($2\sigma$)~\cite{Guo:2017hea}. This result implies that the IH
case in this scenario should be excluded by current observations.
Actually, however, the small upper limit obtained is due mainly to the
strong tension between the Planck data and the latest Hubble
constant measurement. Moreover, in the study in
Ref.~\cite{Guo:2017hea}, the possible mass hierarchies of neutrinos are not
considered. Thus, in the present work, we will revisit the
constraints on the total neutrino mass in the interacting dark
energy (IDE) scenario, and the mass hierarchy cases of neutrinos will be
considered for the first time in this scenario.

The IDE scenario refers to a
cosmological model in which direct exchanges of energy and
momentum between dark energy and dark matter are considered. This
scheme has been proposed and studied widely in the
literature~\cite{Amendola:1999er,Amendola:2001rc,Comelli:2003cv,Cai:2004dk,Zhang:2005rg,Zimdahl:2005bk,Zhang:2004gc,Wang:2006qw,Guo:2007zk,Bertolami:2007zm,Zhang:2007uh,
Boehmer:2008av,Valiviita:2008iv,He:2008tn,He:2009mz,He:2009pd,Koyama:2009gd,Xia:2009zzb,Li:2009zs,Zhang:2009qa,Wei:2010cs,Li:2010ak,He:2010im,Li:2011ga,Fu:2011ab,Zhang:2012uu,Zhang:2013lea,
Li:2013bya,Geng:2015ara,Yin:2015pqa,Murgia:2016ccp,Sola:2016jky,Wang:2016lxa,Pourtsidou:2016ico,Costa:2016tpb,Sola:2016ecz,Feng:2016djj,Xia:2016vnp,vandeBruck:2016hpz,Sola:2016zeg,
Kumar:2017dnp,Sola:2017jbl}. The cosmic coincidence problem can
be greatly alleviated in this
situation~\cite{Comelli:2003cv,Cai:2004dk,Zhang:2005rg,He:2008tn,He:2009pd}.
Searching for interactions between dark energy and dark matter
observationally is an important mission in cosmology. The
impacts of interactions between dark energy and dark matter on
the CMB~\cite{He:2009pd,Pourtsidou:2016ico} and
LSS~\cite{He:2008tn,Amendola:2001rc,Bertolami:2007zm,Koyama:2009gd,Li:2013bya,Pourtsidou:2016ico}
have been investigated in detail.

%Usually, in the Einstein frame, the interaction between the DE scalar field $\phi$ and dark matter can be described by Lagrangian
%\begin{equation}\label{}
%  \mathcal{L}=-\frac{1}{2}\partial^{\mu}\phi \partial_{\mu}\phi-V(\phi)-m(\phi)\overline{\psi}\psi+\mathcal{L}_{\rm kin}[\psi],
%\end{equation}
%where $m(\phi)$ is the mass of matter field $\psi$ and it is a function of the DE scalar field $\phi$. But, actually, owing to the lack of a fundamental theory for dark energy and its possible coupling to dark matter, one has to study this issue from a phenomenological perspective by constructing or parameterizing the forms of the interacting energy transfer rate. Furthermore, we can test the interacting dark energy scenario with the current observational data.

In our work, we investigate a specific class of models of
IDE, in which dark energy is considered to be
the vacuum energy with $w=-1$, and thus the interaction is between
vacuum energy and cold dark matter. In the usual $\Lambda$CDM
model, the vacuum energy density is equivalent to the cosmological
constant $\Lambda$, and in this case, the vacuum energy is a pure
background with a constant $\Lambda$. However, when the
interaction is considered, the vacuum energy density is no longer
a constant, and thus it is no longer a pure background. A model
with such a setting is sometimes called the $\Lambda(t)$CDM model.
In this paper, in order to be consistent with our previous
studies, we call these models I$\Lambda$CDM models.

The energy conservation equations of the vacuum energy density
($\rho_{\Lambda}$) and the cold dark matter density ($\rho_{\rm
c}$) in this scenario are given by
\begin{equation}\label{1.1}
  \dot{\rho}_{\Lambda}=Q,
\end{equation}
\begin{equation}\label{1.2}
  \dot{\rho}_{\rm c}=-3H\rho_{\rm c}-Q,
\end{equation}
where a dot represents the derivative with respect to the cosmic
time $t$, $H$ is the Hubble parameter, and $Q$ is the energy
transfer rate.
%Owing to the lack of a fundamental theory for dark energy and its possible coupling to dark matter, we do not know the concrete form of the energy transfer rate from theory, and thus we can only construct its form from the phenomenological perspective.
In this work, we employ two phenomenological forms of $Q$, i.e.,
$Q=\beta H\rho_{\rm c}$ and $Q=\beta H\rho_{\Lambda}$, where
$\beta$ represents a dimensionless coupling parameter. From
Eqs.~(\ref{1.1}) and~(\ref{1.2}), it can be seen that $\beta>0$
indicates that the energy transport is from dark matter to vacuum
energy, $\beta<0$ represents an inverse energy flow, and $\beta=0$
indicates no interaction between vacuum energy and cold dark matter.

%denotes cold dark matter transferring to vacuum energy, $\beta<0$ denotes vacuum energy transferring to cold dark matter, and $\beta=0$ denotes no interaction between vacuum energy and cold dark matter.

Here we note that the above I$\Lambda$CDM models are based on purely
phenomenological considerations. Because we do not understand the
microscopic mechanism of how dark matter feels a ``fifth force''
through the mediation of dark energy, we cannot describe this
process by a Lagrangian. In fact, we cannot write Lagrangians for most uncoupled dark energy models, let alone for
IDE models. Only for some very specific dark
energy models, e.g., the ``quintessence'' scalar field model, is a
Lagrangian description possible. For the coupled quintessence
model, the Lagrangian in the Einstein frame is given by ${\cal
L}=-\frac{1}{2}\partial^{\mu}\phi
\partial_{\mu}\phi-V(\phi)-m(\phi)\overline{\psi}\psi+\mathcal{L}_{\rm
kin}[\psi]$, where $m(\phi)$ is the mass of the dark matter field
$\psi$, which is a function of the quintessence scalar field $\phi$ in
this scenario. Actually, in this scenario, the forms of the
quintessence potential $V(\phi)$ and the dark matter mass
$m(\phi)$ also need to be assumed. To investigate the
interaction between dark energy and dark matter, it is usually better to consider
more phenomenological scenarios (just as in
the study of dynamical dark energy, some
parametrization models are more apt to be linked to actual
observations). In such a description, an analogy with the
reheating process in the inflationary cosmology or the nuclear
decay process is often made. Namely, one assumes the form of the
energy transfer rate and then writes the energy continuity
equations for dark energy and dark matter. For the I$\Lambda$CDM
models studied in this paper, we actually do not consider the
fundamental theory behind them, but we adopt only a purely
phenomenological perspective. In this scenario, the number of
parameters is the same as that in the $w$CDM cosmology. However, here
we also note that the scenario of a ``running'' vacuum energy
density can actually be related to the renormalization group; see,
e.g., Refs.
\cite{Sola:2016jky,Sola:2016ecz,Sola:2016zeg,Sola:2017jbl}.

Recently, some exciting studies
\cite{Guo:2017hea,Sola:2016jky,Sola:2016ecz,Sola:2016zeg,Sola:2017jbl,Feng:2017usu,Salvatelli:2014zta}
found a nonzero interaction between dark sectors at more
than the $1\sigma$ level. For example, Salvatelli et al.
\cite{Salvatelli:2014zta} showed that a nonzero interaction is
favored at late times. In their work, ten data points from
redshift space distortion (RSD) measurements can break the
degeneracy between $\Omega_{\rm c}h^{2}$ and $\beta$. This can
impose a lower limit on $\Omega_{\rm c}h^{2}$ and lead to a shift
of $\beta$. Actually, when only the RSD data from DR12 are
included, $\beta=0$ is still favored by current observational
data. In addition, in Refs. \cite{Guo:2017hea,Feng:2017usu},
$\beta$ is positively correlated with $\sum m_{\nu}$ in the
$Q=\beta H\rho_{\rm c}$ model. A larger neutrino mass is derived
in this scenario than in the $\Lambda$CDM model. Thus, a
positive value of $\beta$ is favored at more than the $1\sigma$
level. From the above analysis, we conclude that a nonzero
interaction is always dependent on observational data or the model
itself.

In this paper, we report the latest results of constraints on the
total neutrino mass in the I$\Lambda$CDM models. For the neutrino
mass measurement, we consider the NH case, the IH case, and the DH
case. Some important questions will be addressed in
this work: (i) Compared with those in the $\Lambda$CDM model, what
upper bounds of $\sum m_{\nu}$ will be obtained in the
I$\Lambda$CDM models? (ii) Which hierarchy of neutrino masses will
be favored in the I$\Lambda$CDM models? (iii) Can a nonzero
interaction be detected by current cosmological observations?

The structure of the paper is as follows. In
Sec.{~\ref{sec:2}}, we first introduce the observational data
employed in this paper, and then we describe the constraint method
used in our analysis. In Sec.{~\ref{sec:3}}, we analyze the
results of constraining the coupling constant and the neutrino mass
in the I$\Lambda$CDM scenario. The issue of diagnosing the
neutrino mass hierarchy will also be discussed in this section.
Finally, we give the conclusions of the entire work in
Sec.{~\ref{sec:4}}.

\section{Data and method}\label{sec:2}

In what follows, we briefly describe the observational data used
in this work. They are:

\begin{itemize}
\item Planck TT, TE, EE + lowP: We employ the likelihood, including
the TT, EE, and TE spectra, as well as the Planck low-$\ell$
($\ell \leq 30$) likelihood, from the Planck 2015
release~\cite{Ade:2015xua}. \item BAO: We consider the four baryon
acoustic oscillation (BAO) data points, that is, the SDSS-MGS
measurement at $z_{\rm eff}=0.15$~\cite{Ross:2014qpa}, the 6dFGS
measurement at $z_{\rm eff}=0.106$~\cite{Beutler:2011hx}, and the
CMASS and LOWZ samples from the BOSS DR12 at $z_{\rm eff}=0.57$
and $z_{\rm eff}=0.32$~\cite{Anderson:2013zyy}. \item SNIa: We
employ the Joint Light-curve Analysis (JLA) compilation of type
Ia supernovae~\cite{Betoule:2014frx}. It contains 740 type Ia
supernovae data obtained from SNLS and SDSS as well as a few
samples of low-redshift light-curve analysis. \item $H_0$: We use
the new result of direct measurement of the Hubble constant,
$H_{0}=73.00\pm1.75$ km s$^{-1}$ Mpc$^{-1}$~\cite{Riess:2016jrr}.
It reduced the uncertainty from 3.3\% to 2.4\% by using
Wide Field Camera 3 on the Hubble Space Telescope. However,
there is a strong tension between the new $H_{0}$ measurement and
the Planck data. This reminds us to use these data in an
appropriate way. \item RSD: We employ two RSD data points obtained from the LOWZ sample at $z_{\rm
eff}=0.32$ and the CMASS sample at $z_{\rm eff}=0.57$ of the BOSS
DR12~\cite{Gil-Marin:2016wya}. Because these two RSD data points also include the corresponding BAO measurements, we exclude the
BAO measurements of Ref.~\cite{Anderson:2013zyy} from the BAO
likelihood in the combined constraints applied in this paper to
avoid double counting.
    %Note that when the RSD data are considered, the BOSS DR12 results from BAO likelihood will not be included in our calculation.
\end{itemize}

%In addition, other measurements of growth of structure can also affect the constraint on the neutrino mass. For example, the observations of the weak gravitational lensing (WL), the galaxy cluster counts, and the Planck CMB lensing, have been employed to constrain the neutrino mass. However, we do not use these observations in our analysis. The main reasons lie in the following aspects: (i) There are tensions on the fit results of $\sigma_{8}$ between them and the Planck CMB data. (ii) These measurements are analyzed with some uncontrolled systematic effect. (iii) These observations used in this paper are consistent with those of Ref.~\cite{Guo:2017hea}, so that we can compare this work with Ref.~\cite{Guo:2017hea} directly. But, owing to the WL data with a very conservative cut and having a less effect on the neutrino mass, we do not include the WL data in this analysis.

We consider two data combinations in this work, i.e., Planck TT,
TE, EE + lowP + BAO + SNIa + RSD and Planck TT, TE, EE + lowP +
BAO + SNIa + RSD + $H_0$. It should be pointed out that when the
latest local measurement of $H_0$ is included in an analysis,
an additional parameter $N_{\rm eff}$ considered in the
cosmological models will be more helpful for relieving the tension ~\cite{Riess:2016jrr,Zhang:2014ifa,Zhang:2014dxk,DiValentino:2016ucb,Zhang:2017epd,Benetti:2017gvm,Guo:2017qjt,Guo:2018uic}.

Actually, there are also other measurements of the growth of
structure that are often used to constrain the total neutrino
mass, for example, the CMB lensing, galaxy weak lensing, and
cluster counts measurements. In this paper, we consider only the
above two data combinations, mainly because they are convenient for making a direct comparison with the results of
Ref.~\cite{Guo:2017hea}. In addition, the consistency with the
Planck CMB power spectra is also taken into account. In fact, it
is expected that inclusion of the Planck lensing likelihood
would lead to somewhat weaker constraints on the neutrino mass
owing to the low values of $\sigma_8$ preferred by Planck
lensing. Galaxy weak lensing
probes lower redshifts and smaller spatial scales than CMB lensing, and thus the
uncertainties in modeling nonlinearities in the matter power
spectrum and the baryonic feedback on these scales become rather
important. To mitigate the uncertainties of the nonlinear
modeling, a conservative cut scheme can be adopted for the weak
lensing data~\cite{Ade:2015rim}, but this treatment would greatly weaken
the constraining power of the weak lensing data. Moreover, both galaxy weak lensing and cluster
counts actually remain in tension with the Planck CMB data, even though
massive neutrinos are considered in a cosmological model
\cite{Ade:2015xua}. Therefore, these
measurements of structural growth are not used in this paper.

In the I$\Lambda$CDM cosmology, we must consider the
large-scale instability problem~\cite{Valiviita:2008iv} seriously.
To resolve this problem, we adopt the parameterized post-Friedmann
(PPF)
approach~\cite{Guo:2017hea,Li:2014eha,Li:2014cee,Li:2015vla,Zhang:2017ize}.
It is an effective scheme to treat the perturbations of dark
energy. On large scales, a direct relationship is established
between the velocities of dark energy and other components. On
small scales, the Poisson equation can effectively describe
curvature perturbations. In order to be consistent on all scales,
a dynamical function $\Gamma$ is constructed to link them. By
combining the Einstein equations with the conservation equations,
the equation of motion of $\Gamma$ can be determined. Then we can
obtain the correct energy density and velocity perturbations of
dark energy. This PPF scheme can help us explore the entire
parameter space of the I$\Lambda$CDM models without assuming any
specific ranges of $w$ and $\beta$.

For the base $\Lambda$CDM model, the six basic cosmological
parameters are $\{\omega_{\rm b},\, \omega_{\rm c},\,
\theta_{\rm{MC}},\, \tau,\,n_{\rm{s}},\,
{\rm{ln}}(10^{10}A_{\rm{s}})\}$. Here $\omega_{\rm b}=\Omega_{\rm
b} h^2$ is the present density of baryons, and $\omega_{\rm
c}=\Omega_{\rm c} h^2$ is the present density of cold dark matter;
$\theta_{\rm{MC}}$ is the ratio between the sound horizon and the
angular diameter distance at the decoupling epoch; $\tau$ is the
Thomson scattering optical depth resulting from reionization; $n_{\rm{s}}$
is the scalar spectral index; and $A_{\rm{s}}$ is the amplitude of
the primordial power spectrum at the pivot scale, $k_{p}$ = 0.05
Mpc$^{-1}$. For the I$\Lambda$CDM models, the prior range of the
coupling parameter $\beta$ is set to $[-0.2,0.2]$ for the case
of $Q=\beta H\rho_{\rm c}$ and $[-1.0,1.0]$ for the case of
$Q=\beta H\rho_{\Lambda}$. The additional free parameters include
the total neutrino mass $\sum m_\nu$ and the effective number of
relativistic species, $N_{\rm eff}$, with a prior of
$[0,6.0]$.

To constrain the neutrino mass and other cosmological parameters,
we employ a modified version of the publicly available CosmoMC
sampler~\cite{Lewis:2002ah}. The posterior distributions of all
the cosmological parameters can be obtained by fitting to
observational data.

For the NH case, the neutrino mass spectrum is
\begin{equation}\label{2.1}
  (m_{1},m_{2},m_{3})=(m_{1},\sqrt{m_{1}^{2}+\Delta m_{21}^{2}},\sqrt{m_{1}^{2}+|\Delta m_{31}^{2}|})
\end{equation}
in terms of a free parameter $m_{1}$. For the IH case, the
neutrino mass spectrum is
\begin{equation}\label{2.2}
\small (m_{1},m_{2},m_{3})=(\sqrt{m_{3}^{2}+|\Delta
m_{31}^{2}|},\sqrt{m_{3}^{2}+|\Delta m_{31}^{2}|+\Delta
m_{21}^{2}},m_{3})
\end{equation}
in terms of a free parameter $m_{3}$. We also consider the DH case
for comparison. In this case, the neutrino mass spectrum is
\begin{equation}\label{2.3}
  m_{1}=m_{2}=m_{3}=m,
\end{equation}
where $m$ is a free parameter. It should be pointed out that the
input lower bounds of $\sum m_\nu$ are 0.06 eV for the NH case,
0.10 eV for the IH case, and 0 eV for the DH case.

\section{results}\label{sec:3}

We constrain the total neutrino mass in the I$\Lambda$CDM models.
For comparison with Ref.~\cite{Guo:2017hea}, we further consider the
three mass hierarchy cases of neutrinos, i.e., the NH case, the IH
case, and the DH case. Our fitting results are listed in
Tables~\ref{tab:l} and \ref{tab:2} for the base $\Lambda$CDM model
and Tables~\ref{tab:3}--\ref{tab:6} for the two I$\Lambda$CDM
models. For convenient display, we use ``I$\Lambda$CDM1'' and
``I$\Lambda$CDM2'' instead of ``the $Q\propto \rho_{\rm c}$ model'' and
``the $Q\propto \rho_{\Lambda}$ model,'' respectively. In these
tables, we quote the $\pm1\sigma$ errors of cosmological
parameters, but only the $2\sigma$ upper limit is given for the total
neutrino mass $\sum m_\nu$. In addition, we also list the values
of $\chi^{2}_{\rm min}$.

\subsection{Neutrino mass}

\begin{table*}[!htp]
\caption{Fitting results of the cosmological parameters in the
$\Lambda$CDM+$\sum m_\nu$ model for mass hierarchy cases
NH, IH, and DH.} \centering
\renewcommand{\arraystretch}{1.3}
\scalebox{1}[1]{%
\begin{tabular}{|c|c c c|}
\hline
\multicolumn{1}{|c|}{Data}&\multicolumn{3}{c|}{Planck TT,TE,EE+lowP+BAO+SNIa+RSD}     \\
 \cline{1-1}\cline{2-4}
 Model&$\Lambda$CDM+$\sum m_{\nu}^{\rm NH}$&$\Lambda$CDM+$\sum m_{\nu}^{\rm IH}$&$\Lambda$CDM+$\sum m_{\nu}^{\rm DH}$\\
\hline $\Omega_bh^2$    &$0 .02231\pm0.00014$
                 &$0 .02232\pm0.00014$
                 &$0 .02230\pm0.00014$\\

$\Omega_ch^2$    &$0 .1186\pm0.0011$
                 &$0 .1183\pm0.0011$
                 &$0 .1188\pm0.0012$ \\

$100\theta_{\emph{\rm MC}}$     &$1 .04087\pm0.00029$
                                &$1 .04087\pm0.00030$
                                &$1 .04084\pm0.00029$  \\

$\tau$                  &$0 .078^{+0.017}_{-0.016}$
                        &$0 .082^{+0.017}_{-0.016}$
                        &$0 .074\pm0.017$\\

${\textrm{ln}}(10^{10}A_s)$      &$3 .087\pm0.032$
                                 &$3 .094\pm0.032$
                                 &$3 .080\pm0.033$ \\

$n_s$             &$0 .9672\pm0.0042$
                  &$0 .9678\pm0.0041$
                  &$0 .9665\pm0.0042$ \\

\hline

$\sum m_\nu$       &$<0.217~\textrm{eV}$
                   &    $<0.235~\textrm{eV}$
                   &    $<0.198~\textrm{eV}$   \\

\hline

$\Omega_{\rm m}$     &$0 .3139^{+0.0071}_{-0.0072}$
                     &$0 .3159^{+0.0069}_{-0.0070}$
                     &$0 .3116^{+0.0072}_{-0.0081}$\\

$H_0$                 &$67 .31^{+0.59}_{-0.54}$
                      &$67 .13^{+0.53}_{-0.52}$
                      &$67 .53^{+0.64}_{-0.57}$ \\

\hline $\chi^2_{\rm min}$     &$13661.472$
                       &$13661.662$
                       &$13658.622$ \\
\hline
\end{tabular}}
\label{tab:l}
\end{table*}

\begin{table*}[!htp]
\caption{Fitting results of the cosmological parameters in the
$\Lambda$CDM+$\sum m_\nu$+$N_{\rm eff}$ model for mass
hierarchy cases NH, IH, and DH.} \centering
\renewcommand{\arraystretch}{1.3}
\scalebox{1}[1]{%
\begin{tabular}{|c|c c c|}
\hline
\multicolumn{1}{|c|}{Data}&\multicolumn{3}{c|}{Planck TT,TE,EE+lowP+BAO+SNIa+RSD+$H_0$}     \\
\cline{1-1}\cline{2-4}
Model&$\Lambda$CDM+$\sum m_{\nu}^{\rm NH}$+$N_{\rm eff}$&$\Lambda$CDM+$\sum m_{\nu}^{\rm IH}$+$N_{\rm eff}$&$\Lambda$CDM+$\sum m_{\nu}^{\rm DH}$+$N_{\rm eff}$\\
\hline $\Omega_bh^2$     &$0 .02252^{+0.00017}_{-0.00018}$
                  &$0 .02256\pm0.00017$
                  &$0 .02248\pm0.00018$ \\

$\Omega_ch^2$     &$0 .1214\pm0.0028$
                  &$0 .1217\pm0.0027$
                  &$0 .1211^{+0.0027}_{-0.0028}$ \\

$100\theta_{\emph{\rm MC}}$          &$1 .04058\pm0.00041$
                                     &$1 .04054\pm0.00040$
                                     &$1 .04063\pm0.00040$  \\

$\tau$                 &$0 .082\pm0.017$
                       &$0 .086\pm0.017$
                       &$0 .077\pm0.017$\\

${\textrm{ln}}(10^{10}A_s)$      &$3 .103\pm0.034$
                                 &$3 .112\pm0.034$
                                 &$3 .093\pm0.035$\\

$n_s$             &$0 .9761^{+0.0067}_{-0.0068}$
                  &$0 .9778^{+0.0066}_{-0.0067}$
                  &$0 .9740^{+0.0068}_{-0.0073}$\\

\hline

$\sum m_\nu$             &$     <0.227~\textrm{eV}$
                         &  $<0.249~\textrm{eV}$
                         &  $<0.202~\textrm{eV}$   \\

$N_{\rm eff}$        &$3 .27\pm0.16$
                     &$3 .30\pm0.16$
                     &$3 .23\pm0.16$  \\

\hline $\Omega_{\rm m}$                &$0
.3058^{+0.0068}_{-0.0067}$
                                &$0 .3071\pm0.0067$
                                &$0 .3040^{+0.0069}_{-0.0068}$\\

$H_0$            &$68 .93\pm0.97$
                 &$68 .94\pm0.97$
                 &$68 .94\pm0.98$ \\

\hline $\chi^2_{\rm min}$         &        $13669.956$
                           &        $13670.988$
                           &        $13668.192$ \\
\hline
\end{tabular}}
\label{tab:2}
\end{table*}

\begin{table*}[!htp]
\caption{Fitting results of the cosmological parameters in the
I$\Lambda$CDM1 ($Q=\beta H \rho_{\rm c}$)+$\sum m_\nu$ model for mass hierarchy cases NH, IH, and DH.}
\centering
\renewcommand{\arraystretch}{1.3}
\scalebox{1}[1]{%
\begin{tabular}{|c|c c c|}
\hline
\multicolumn{1}{|c|}{Data}&\multicolumn{3}{c|}{Planck TT,TE,EE+lowP+BAO+SNIa+RSD}     \\
\cline{1-1}\cline{2-4}
Model&I$\Lambda$CDM1+$\sum m_{\nu}^{\rm NH}$&I$\Lambda$CDM1+$\sum m_{\nu}^{\rm IH}$&I$\Lambda$CDM1+$\sum m_{\nu}^{\rm DH}$\\
\hline $\Omega_bh^2$    &$0 .02228\pm0.00015$
                 &$0 .02227\pm0.00016$
                 &$0 .02229\pm0.00015$\\

$\Omega_ch^2$    &$0 .1182^{+0.0014}_{-0.0013}$
                 &$0 .1178^{+0.0015}_{-0.0013}$
                 &$0 .1187^{+0.0015}_{-0.0013}$ \\

$100\theta_{\emph{\rm MC}}$     &$1 .04087\pm0.00030$
                                &$1 .04088\pm0.00030$
                                &$1 .04085\pm0.00030$  \\

$\tau$                  &$0 .078\pm0.017$
                        &$0 .081\pm0.017$
                        &$0 .074\pm0.017$\\

${\textrm{ln}}(10^{10}A_s)$      &$3 .089\pm0.033$
                                 &$3 .095^{+0.033}_{-0.032}$
                                 &$3 .081\pm0.033$ \\

$n_s$             &$0 .9667\pm0.0043$
                  &$0 .9669\pm0.0044$
                  &$0 .9663\pm0.0044$ \\

\hline

$\beta$      &$0 .0010\pm0.0018$
             &$0 .0014^{+0.0017}_{-0.0018}$
             &$0 .0004\pm0.0018$  \\

$\sum m_\nu$       &$<0.279~\textrm{eV}$
                   &$<0.301~\textrm{eV}$
                   &$<0.245~\textrm{eV}$   \\

\hline

$\Omega_{\rm m}$     &$0 .3116\pm0.0090$
                     &$0 .3121\pm0.0090$
                     &$0 .3112^{+0.0090}_{-0.0089}$\\

$H_0$                 &$67 .52\pm0.74$
                      &$67 .47\pm0.74$
                      &$67 .57^{+0.73}_{-0.74}$\\

\hline $\chi^2_{\rm min}$     &$13660.994$
                       &$13663.684$
                       &$13661.990$ \\
\hline
\end{tabular}}
\label{tab:3}
\end{table*}

\begin{table*}[!htp]
\caption{Fitting results of the cosmological parameters in the
I$\Lambda$CDM1 ($Q=\beta H \rho_{\rm c}$)+$\sum m_\nu$+$N_{\rm
eff}$ model for mass hierarchy cases NH, IH, and DH.} \centering
\renewcommand{\arraystretch}{1.3}
\scalebox{1}[1]{%
\begin{tabular}{|c|c c c|}
\hline
\multicolumn{1}{|c|}{Data}&\multicolumn{3}{c|}{Planck TT,TE,EE+lowP+BAO+SNIa+RSD+$H_0$}     \\
\cline{1-1}\cline{2-4}
Model&I$\Lambda$CDM1+$\sum m_{\nu}^{\rm NH}$+$N_{\rm eff}$&I$\Lambda$CDM1+$\sum m_{\nu}^{\rm IH}$+$N_{\rm eff}$&I$\Lambda$CDM1+$\sum m_{\nu}^{\rm DH}$+$N_{\rm eff}$\\
\hline $\Omega_bh^2$     &$0 .02244\pm0.00021$
                  &$0 .02244\pm0.00021$
                  &$0 .02242\pm0.00021$\\

$\Omega_ch^2$     &$0 .1201\pm0.0032$
                  &$0 .1200\pm0.0032$
                  &$0 .1203\pm0.0032$ \\

$100\theta_{\emph{\rm MC}}$          &$1 .04067\pm0.00042$
                                     &$1 .04064\pm0.00042$
                                     &$1 .04068\pm0.00042$  \\

$\tau$                 &$0 .081^{+0.018}_{-0.017}$
                       &$0 .084\pm0.018$
                       &$0 .077\pm0.017$\\

${\textrm{ln}}(10^{10}A_s)$      &$3 .102^{+0.036}_{-0.035}$
                                 &$3 .109^{+0.036}_{-0.035}$
                                 &$3 .093\pm0.035$\\

$n_s$            &$0 .9738\pm0.0078$
                 &$0 .9744^{+0.0078}_{-0.0077}$
                 &$0 .9724^{+0.0078}_{-0.0077}$\\

\hline

$\beta$               &$0 .0015\pm0.0019$
                      &$0 .0018\pm0.0019$
                      &$0 .0010^{+0.0019}_{-0.0021}$   \\

$\sum m_\nu$             &$     <0.296~\textrm{eV}$
                         &  $<0.321~\textrm{eV}$
                         &  $<0.263~\textrm{eV}$   \\

$N_{\rm eff}$        &$3 .22\pm0.18$
                     &$3 .23\pm0.17$
                     &$3 .20\pm0.17$  \\

\hline $\Omega_{\rm m}$                &$0
.3024^{+0.0081}_{-0.0080}$
                                &$0 .3029\pm0.0082$
                                &$0 .3021\pm0.0082$\\

$H_0$            &$69 .05^{+0.99}_{-0.98}$
                 &$69 .05\pm0.99$
                 &$69 .00\pm^{+0.98}_{-0.99}$ \\

\hline $\chi^2_{\rm min}$         &        $13668.898$
                           &            $13671.246$
                           &            $13668.984$ \\
\hline
\end{tabular}}
\label{tab:4}
\end{table*}

\begin{table*}[!htp]
\caption{Fitting results of the cosmological parameters in the
I$\Lambda$CDM2 ($Q=\beta H \rho_{\Lambda}$)+$\sum m_\nu$ model
for mass hierarchy cases NH, IH, and DH.}
\centering
\renewcommand{\arraystretch}{1.3}
\scalebox{1}[1]{%
\begin{tabular}{|c|c c c|}
\cline{1-1}\cline{2-4}
\multicolumn{1}{|c|}{Data}&\multicolumn{3}{c|}{Planck TT,TE,EE+lowP+BAO+SNIa+RSD}     \\
\cline{1-1}\cline{2-4}
Model&I$\Lambda$CDM2+$\sum m_{\nu}^{\rm NH}$&I$\Lambda$CDM2+$\sum m_{\nu}^{\rm IH}$&I$\Lambda$CDM2+$\sum m_{\nu}^{\rm DH}$\\
\hline $\Omega_bh^2$    &$0 .02233\pm0.00014$
                 &$0 .02235\pm0.00014$
                 &$0 .02232\pm0.00014$\\

$\Omega_ch^2$    &$0 .1319^{+0.0150}_{-0.0068}$
                 &$0 .1311^{+0.0153}_{-0.0069}$
                 &$0 .1335^{+0.0146}_{-0.0064}$\\

$100\theta_{\emph{\rm MC}}$     &$1 .04017^{+0.00049}_{-0.00083}$
                                &$1 .04022^{+0.00049}_{-0.00082}$
                                &$1 .04009^{+0.00046}_{-0.00080}$  \\

$\tau$                  &$0 .079\pm0.016$
                        &$0 .082\pm0.016$
                        &$0 .074^{+0.017}_{-0.016}$\\

${\textrm{ln}}(10^{10}A_s)$      &$3 .089\pm0.032$
                                 &$3 .095\pm0.032$
                                 &$3 .080\pm0.032$ \\

$n_s$             &$0 .9678^{+0.0040}_{-0.0041}$
                  &$0 .9685\pm0.0041$
                  &$0 .9670\pm0.0042$ \\

\hline

$\beta$      &$-0.129^{+0.066}_{-0.146}$
             &$-0.125^{+0.067}_{-0.151}$
             &$-0.140^{+0.060}_{-0.141}$  \\

$\sum m_\nu$       &$<0.206~\textrm{eV}$
                   &$<0.228~\textrm{eV}$
                   &$<0.180~\textrm{eV}$   \\

\hline

$\Omega_{\rm m}$     &$0 .343^{+0.033}_{-0.018}$
                     &$0 .344^{+0.033}_{-0.018}$
                     &$0 .343^{+0.032}_{-0.017}$\\

$H_0$                 &$67 .32^{+0.54}_{-0.53}$
                      &$67 .15\pm0.52$
                      &$67 .58^{+0.61}_{-0.56}$\\

\hline $\chi^2_{\rm min}$     &$13658.662$
                       &$13659.402$
                       &$13656.462$ \\
\hline
\end{tabular}}
\label{tab:5}
\end{table*}

\begin{table*}[!htp]
\caption{Fitting results of the cosmological parameters in the
I$\Lambda$CDM2 ($Q=\beta H \rho_{\Lambda}$)+$\sum m_\nu$+$N_{\rm
eff}$ model for mass hierarchy cases NH, IH, and DH.} \centering
\renewcommand{\arraystretch}{1.3}
\scalebox{1}[1]{%
\begin{tabular}{|c|c c c|}
\hline
\multicolumn{1}{|c|}{Data}&\multicolumn{3}{c|}{Planck TT,TE,EE+lowP+BAO+SNIa+RSD+$H_0$}     \\
\cline{1-1}\cline{2-4}
Model&I$\Lambda$CDM2+$\sum m_{\nu}^{\rm NH}$+$N_{\rm eff}$&I$\Lambda$CDM2+$\sum m_{\nu}^{\rm IH}$+$N_{\rm eff}$&I$\Lambda$CDM2+$\sum m_{\nu}^{\rm DH}$+$N_{\rm eff}$\\
\hline $\Omega_bh^2$     &$0 .02259\pm0.00018$
                  &$0 .02264\pm0.00018$
                  &$0 .02255\pm0.00018$\\

$\Omega_ch^2$     &$0 .1415^{+0.0144}_{-0.0076}$
                  &$0 .1417^{+0.0144}_{-0.0079}$
                  &$0 .1415^{+0.0139}_{-0.0076}$ \\

$100\theta_{\emph{\rm MC}}$          &$1
.03950^{+0.00057}_{-0.00085}$
                                     &$1 .03947^{+0.00058}_{-0.00084}$
                                     &$1 .03953^{+0.00056}_{-0.00083}$  \\

$\tau$                 &$0 .084\pm0.017$
                       &$0 .088^{+0.016}_{-0.017}$
                       &$0 .079^{+0.017}_{-0.018}$\\

${\textrm{ln}}(10^{10}A_s)$      &$3 .109\pm0.034$
                                 &$3 .118\pm0.033$
                                 &$3 .098\pm0.034$\\

$n_s$            &$0 .9792\pm0.0070$
                 &$0 .9811\pm0.0070$
                 &$0 .9769\pm0.0071$\\

\hline

$\beta$               &$-0.172^{+0.050}_{-0.122}$
                      &$-0.171^{+0.052}_{-0.122}$
                      &$-0.174^{+0.049}_{-0.118}$   \\

$\sum m_\nu$             &  $<0.215~\textrm{eV}$
                         &  $<0.245~\textrm{eV}$
                         &  $<0.188~\textrm{eV}$   \\

$N_{\rm eff}$        &$3 .33^{+0.16}_{-0.17}$
                     &$3 .37\pm0.17$
                     &$3 .29\pm0.17$ \\

\hline $\Omega_{\rm m}$                &$0 .344^{+0.028}_{-0.014}$
                                &$0 .345^{+0.027}_{-0.014}$
                                &$0 .343^{+0.027}_{-0.014}$\\

$H_0$            &$69 .30\pm1.00$
                 &$69 .30\pm1.00$
                 &$69 .30\pm1.00$ \\

\hline $\chi^2_{\rm min}$         &        $13664.296$
                           &            $13665.398$
                           &            $13664.704$ \\
\hline
\end{tabular}}
\label{tab:6}
\end{table*}

We first use the Planck TT, TE, EE + lowP + BAO + SNIa + RSD data
combination to constrain these models. In the $\Lambda$CDM+$\sum
m_\nu$ model, we obtain $\sum m_\nu<0.217$ eV for the NH case,
$\sum m_\nu<0.235$ eV for the IH case, and $\sum m_\nu<0.198$ eV
for the DH case (see Table~\ref{tab:l}). In the
I$\Lambda$CDM1+$\sum m_\nu$ model, the constraint results become
$\sum m_\nu<0.279$ eV for the NH case, $\sum m_\nu<0.301$ eV for
the IH case, and $\sum m_\nu<0.245$ eV for the DH case (see
Table~\ref{tab:3}), indicating that much looser constraints are
obtained than those in the $\Lambda$CDM+$\sum m_\nu$ model. Further,
in the I$\Lambda$CDM2+$\sum m_\nu$ model, the results are $\sum
m_\nu<0.206$ eV for the NH case, $\sum m_\nu<0.228$ eV for the IH
case, and $\sum m_\nu<0.180$ eV for the DH case (see
Table~\ref{tab:5}), indicating that the constraints are slightly tighter than those in the $\Lambda$CDM+$\sum m_\nu$ model.

Furthermore, we consider including the latest local measurement
of the Hubble constant, $H_{0}=73.00\pm1.75$ km s$^{-1}$
Mpc$^{-1}$, to constrain these models. Using the Planck TT, TE,
EE + lowP + BAO + SNIa + RSD + $H_0$ data combination, we find
that, compared with the $\Lambda$CDM+$\sum m_\nu$+$N_{\rm eff}$
model, the I$\Lambda$CDM1+$\sum m_\nu$+$N_{\rm eff}$ model
provides much looser constraints on $\sum m_\nu$, whereas the
I$\Lambda$CDM2+$\sum m_\nu$+$N_{\rm eff}$ model provides tighter
constraints. This is consistent with the case using the Planck TT,
TE, EE + lowP + BAO + SNIa + RSD data. The detailed results are given in Tables~\ref{tab:2},~\ref{tab:4}, and~\ref{tab:6}.
Considering the three mass hierarchies, we find that the value of
$\sum m_\nu$ is smallest in the DH case and largest in the
IH case. Thus, the mass hierarchy can affect the constrained
values of $\sum m_\nu$ in these models.

Next, we give the values of $\chi^2_{\rm min}$ for the three mass
hierarchy cases. For the $\Lambda$CDM+$\sum m_\nu$ model and the
$\Lambda$CDM+$\sum m_\nu$+$N_{\rm eff}$ model, we find that the
values of $\chi^2_{\rm min}$ in the NH case are slightly smaller
than those in the IH case, and the difference $\Delta \chi^2_{\rm
min} \equiv \chi^2_{\rm IH; min}-\chi^2_{\rm NH; min}<2$ (see
Tables~\ref{tab:l} and~\ref{tab:2}). When the mass splittings are
neglected, the values of $\chi^2_{\rm min}$ are smallest in
this case. In addition, for the I$\Lambda$CDM1+$\sum m_\nu$ model and the
I$\Lambda$CDM1+$\sum m_\nu$+$N_{\rm eff}$ model, we find that the
difference $\Delta \chi^2_{\rm min} \equiv \chi^2_{\rm IH;
min}-\chi^2_{\rm NH; min}>2$ (see Tables~\ref{tab:3}
and~\ref{tab:4}), further providing strong support for the NH
case. In this situation, the values of $\chi^2_{\rm min}$ for the
DH case are the smallest. For the I$\Lambda$CDM2+$\sum m_\nu$
model and the I$\Lambda$CDM2+$\sum m_\nu$+$N_{\rm eff}$ model, we
obtain results consistent with those in the $\Lambda$CDM+$\sum
m_\nu$ model and the $\Lambda$CDM+$\sum m_\nu$+$N_{\rm eff}$
model. Namely, the NH case is favored over the IH case, but
the difference $\Delta \chi^2_{\rm min} \equiv \chi^2_{\rm IH;
min}-\chi^2_{\rm NH; min}<2$ (see Tables~\ref{tab:5}
and~\ref{tab:6}), which does not seem to be significant enough to
distinguish between the mass hierarchies.

\subsection{Coupling parameter}

\begin{figure*}[ht!]
\begin{center}
\includegraphics[width=7cm]{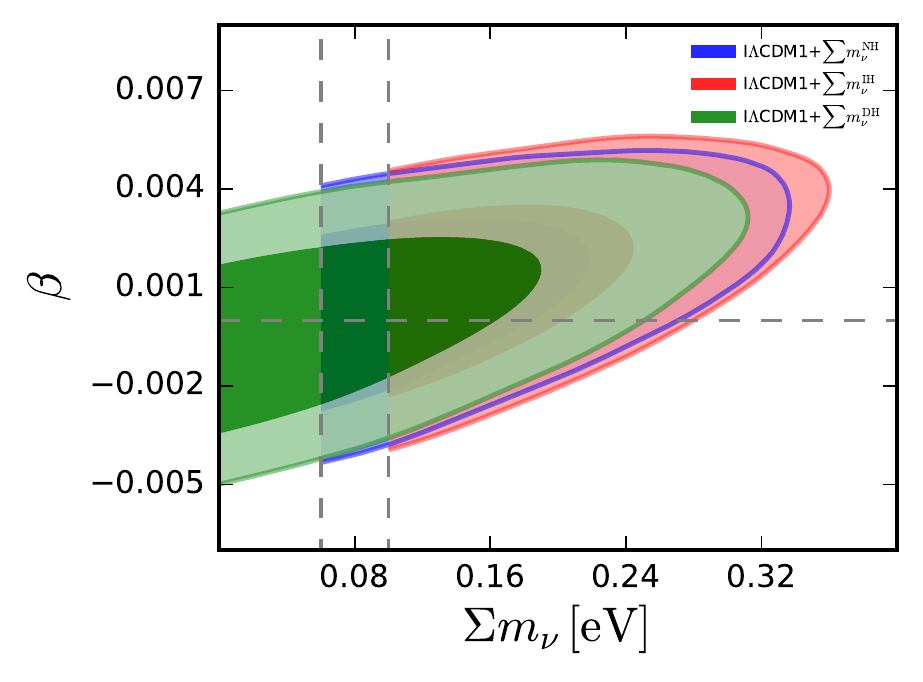}
\end{center}
\caption{Two-dimensional marginalized contours ($1\sigma$ and
$2\sigma$) of the $\sum m_\nu$--$\beta$ plane in the
I$\Lambda$CDM1 ($Q=\beta H \rho_{\rm c}$)+$\sum m_\nu$ model
using the Planck TT, TE, EE + lowP + BAO + SNIa + RSD data combination for various neutrino mass hierarchies.}
\label{fig1}
\end{figure*}

\begin{figure*}[ht!]
\begin{center}
\includegraphics[width=7cm]{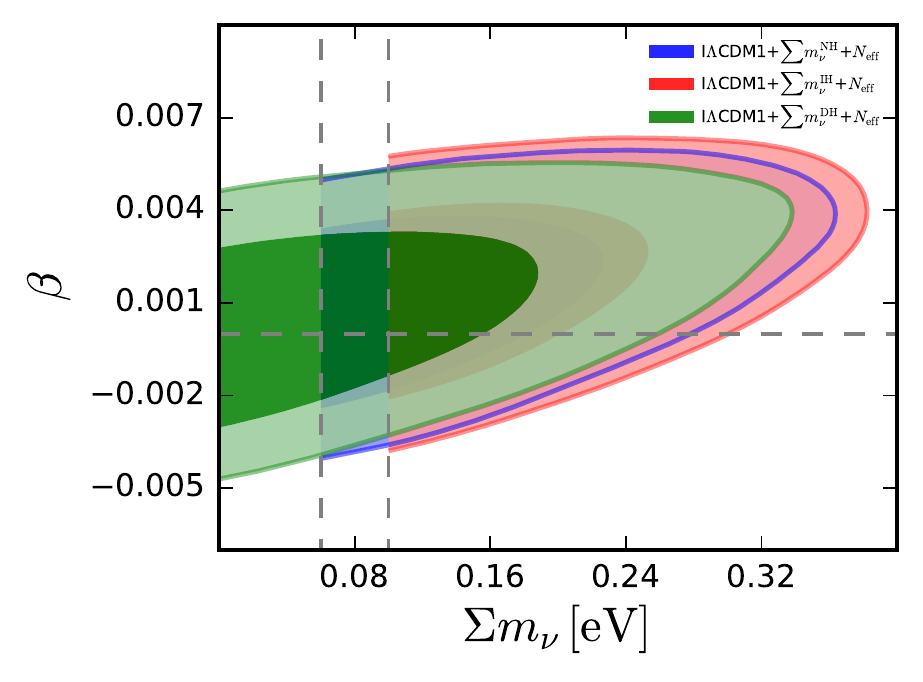}
\end{center}
\caption{Two-dimensional marginalized contours ($1\sigma$ and
$2\sigma$) of the $\sum m_\nu$--$\beta$ plane in the
I$\Lambda$CDM1 ($Q=\beta H \rho_{\rm c}$)+$\sum m_\nu$+$N_{\rm
eff}$ model using the Planck TT, TE, EE + lowP + BAO + SNIa + RSD +$H_0$
data combination for various neutrino mass
hierarchies.}
\label{fig2}
\end{figure*}

\begin{figure*}[ht!]
\begin{center}
\includegraphics[width=9cm]{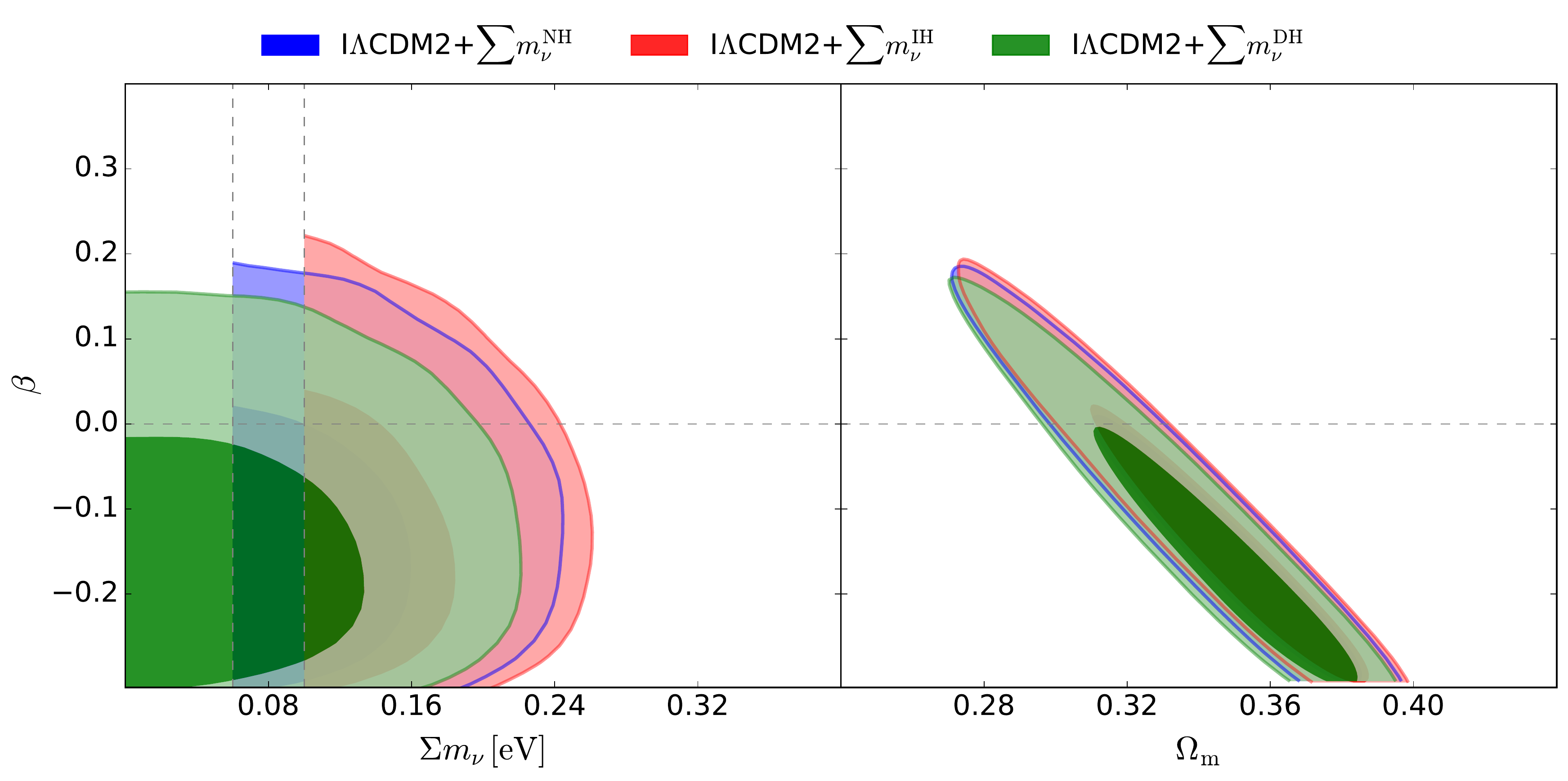}
\end{center}
\caption{Two-dimensional marginalized contours ($1\sigma$ and
$2\sigma$) of the $\sum m_\nu$--$\beta$ plane and the $\Omega_{\rm
m}$--$\beta$ plane in the I$\Lambda$CDM2 ($Q=\beta H
\rho_{\Lambda}$)+$\sum m_\nu$ model using the Planck
TT, TE, EE + lowP + BAO + SNIa + RSD data combination for various neutrino mass hierarchies.}
\label{fig3}
\end{figure*}

\begin{figure*}[ht!]
\begin{center}
\includegraphics[width=9cm]{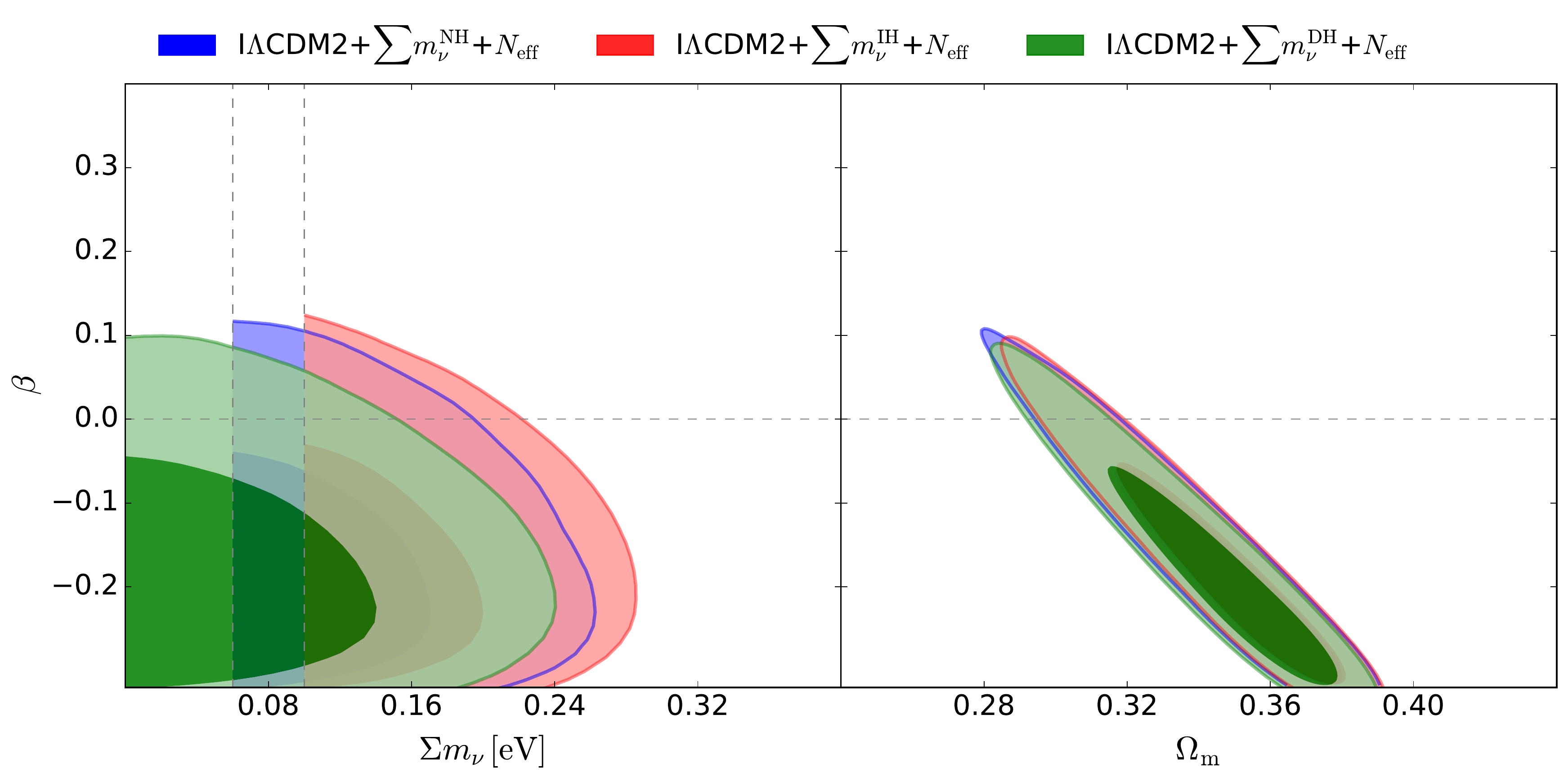}
\end{center}
\caption{Two-dimensional marginalized contours ($1\sigma$ and
$2\sigma$) of the $\sum m_\nu$--$\beta$ plane and the $\Omega_{\rm
m}$--$\beta$ plane in the I$\Lambda$CDM2 ($Q=\beta H
\rho_{\Lambda}$)+$\sum m_\nu$+$N_{\rm eff}$ model using the
Planck TT, TE, EE +lowP +BAO + SNIa + RSD + $H_0$ data combination for various neutrino mass hierarchies.}
\label{fig4}
\end{figure*}

In this subsection, we discuss the fitting results of the coupling
parameter $\beta$. First, we constrain the I$\Lambda$CDM1+$\sum
m_\nu$ model using the Planck TT, TE, EE + lowP + BAO + SNIa +
RSD data. The detailed fitting results are given in
Table~\ref{tab:3}. We see that $\beta=0$ is favored within the
$1\sigma$ range, regardless of the neutrino mass hierarchy. Furthermore, we constrain the I$\Lambda$CDM1+$\sum
m_\nu$+$N_{\rm eff}$ model using the Planck TT, TE, EE + lowP +
BAO + SNIa + RSD + $H_0$ data. The detailed fitting results are
given in Table~\ref{tab:4}. For this data combination,
$\beta=0$ is still favored by the data. Thus, there
is no evidence of a nonzero interaction in the $Q=\beta H\rho_{\rm
c}$ model. In addition, from Figs.~\ref{fig1} and~\ref{fig2}, we
see that $\beta$ is positively correlated with $\sum m_\nu$.

Next, we constrain the I$\Lambda$CDM2+$\sum m_\nu$ model using
the Planck TT, TE, EE + lowP + BAO + SNIa + RSD data, and we
constrain the I$\Lambda$CDM2+$\sum m_\nu$+$N_{\rm eff}$ model
using the Planck TT, TE, EE + lowP + BAO + SNIa + RSD + $H_0$
data. The detailed fitting results are given in Tables~\ref{tab:5}
and~\ref{tab:6}, respectively. For the $Q=\beta H\rho_{\rm \Lambda}$ model, an
exciting result is that negative values of $\beta$ are favored
by current observations at more than the $1\sigma$ level,
indicating that vacuum energy decays into cold dark matter.
Further, we see that the values of $\beta$ are truncated when
$\beta<-0.3$. This is because $\beta$ is anticorrelated with
$\Omega_{\rm m}$, as shown in Figs.~\ref{fig3} and~\ref{fig4}. A
larger $\Omega_{\rm m}$ leads to a smaller $\beta$, whereas a too-small value of $\beta$ (negative value) is not allowed by theory
in current cosmology.

\section{Conclusion}\label{sec:4}

In this paper, we constrain the total neutrino mass in the
scenario of vacuum energy interacting with cold dark matter. We
consider three neutrino mass hierarchy cases, i.e., the NH case,
the IH case, and the DH case. In our analysis, we employ two data
combinations, i.e., the Planck TT, TE, EE + lowP + BAO + SNIa +
RSD data combination and the Planck TT, TE, EE + lowP + BAO + SNIa
+ RSD + $H_0$ data combination. It is worth mentioning that there
is a strong tension between the local measurement of $H_0$ and the
fitting result derived from the Planck data. Thus, when the local
measurement of $H_0$ is used to constrain the models, we consider
an additional parameter, $N_{\rm eff}$, in the cosmological models
to relieve the tension.

We find that, compared with the $\Lambda$CDM model, the $Q=\beta H
\rho_{\rm c}$ model can provide a much looser constraint on the
total neutrino mass, whereas the $Q=\beta H \rho_{\Lambda}$ model
gives a slightly tighter constraint. We also compare the
constrained values of $\sum m_\nu$ for the three mass
hierarchy cases. We find that the upper limits on $\sum m_\nu$ are
smallest in the DH case. By comparing the values of $\chi^2_{\rm
min}$ for different neutrino mass hierarchies, we find that the NH
case is favored over the IH case in the I$\Lambda$CDM models.
The difference $\Delta \chi^2_{\rm min} \equiv \chi^2_{\rm IH;
min}-\chi^2_{\rm NH; min}$ is 2.69 in the $Q=\beta H
\rho_{\rm c}$ model. Our results are consistent with those of
Refs.~\cite{Huang:2015wrx,Wang:2016tsz,Yang:2017amu}.

In addition, we also probe the interaction between vacuum energy
and cold dark matter. For the $Q=\beta H \rho_{\rm c}$ model,
there is no evidence of a nonzero interaction. However, for the
$Q=\beta H \rho_{\Lambda}$ model, we find that a negative $\beta$
is favored at more than the $1\sigma$ level, indicating that
vacuum energy decays into cold dark matter. Our fitting results of
the coupling constant $\beta$ are different from those of
Ref.~\cite{Guo:2017hea}. The reason may be that the
local $H_0$ measurement with a strong tension is employed in
Ref.~\cite{Guo:2017hea}, whereas in our analysis, we add the
parameter $N_{\rm eff}$ to alleviate the tension when the local
measurement value of $H_0$ is employed to constrain the models.

\begin{acknowledgments}

This work was supported by the National Natural Science Foundation of China (Grants No.~11522540 and No.~11690021), the Top-Notch Young Talents Program of China, and the Provincial Department of Education of Liaoning (Grant No.~ L2012087).

\end{acknowledgments}

\end{document}